\documentclass[usenatbib]{mn2e}
\usepackage[dvips]{graphicx,color}
\usepackage{amssymb}
\title[Very long-term X-ray variations in LMXBs: solar cycle-like variations in the donor?]
  {Very long-term X-ray variations in LMXBs: solar cycle-like variations in the donor?}
\author[M.M. Kotze et al.]
  {M.M.~Kotze $^{1,2}$,
  P.A.~Charles $^{1,3}$\\
  $^1$ South African Astronomical Observatory,
       P.O. Box 9, Observatory 7935, South Africa\\
  $^2$ Dept. of Astronomy, University of Cape Town, South Africa\\
  $^3$ School of Physics \& Astronomy, University of Southampton, Southampton SO17 1BJ, UK \\
}
\date{Released 2009 }

\pagerange{\pageref{firstpage}--\pageref{lastpage}} \pubyear{2009}

\def\LaTeX{L\kern-.36em\raise.3ex\hbox{a}\kern-.15em
    T\kern-.1667em\lower.7ex\hbox{E}\kern-.125emX}

\def\deg{$^{\circ}$}

\begin{document}

\label{firstpage}

\maketitle

\begin{abstract}
\noindent Long-term monitoring of Low Mass X-ray Binaries (LMXBs) by the All Sky Monitor on board the Rossi X-ray Timing Explorer now covers $\sim$13 yrs and shows that certain LMXB types display very long-term ($\sim$ several to tens of years) quasi-periodic modulations. These timescales are much longer than any ``super-orbital'' periods reported hitherto and likely have a different origin. We suggest here that they are due to long-term variations in the mass-transfer rate from the donor, which are a consequence of solar-like magnetic cycles that lead to $P_{orb}$ changes (as proposed by Richman, Applegate \& Patterson 1994 for similar long-term variations in CVs). Atoll sources display much larger amplitude modulations than Z sources over these timescales, presumably because Z sources are Eddington limited and hence unable to respond as readily as Atoll sources to fluctuations in the mass-transfer rate from the donor.

\end{abstract}

\section{Introduction} 

Low Mass X-ray Binaries (LMXBs) contain a neutron star (NS) or black hole (BH) primary onto which material is transferred from a low-mass ($M \leq 1 M_{\odot}$) late-type, essentially normal main sequence (MS) star (spectral type $\sim$A-M), with longer period systems containing a sub-giant. There are $\sim$190 luminous ($\lesssim10^{38}$ erg\space s$^{-1}$) LMXBs known in our galaxy (Liu, van Paradijs \& van den Heuvel 2007). Cataclysmic Variables (CVs) are very similar, but have white dwarf (WD) primaries at a $\sim10^3$ factor reduction in luminosity and have orbital periods that range from hours to $\sim10$ d. While the flux from LMXBs is predominantly in X-rays which originate from the inner accretion disc and the NS surface (where applicable), the flux from CVs is predominantly in the optical and originates from the entire accretion disc, the hot spot(s) on the WD surface and the disc-mass-transfer stream impact region. (For details see e.g. Frank, King \& Raine 1992).

Long-term, non-orbital, quasi-periodic variations on timescales of $\sim$tens-hundreds of days in LMXBs were discovered early in X-ray astronomy (see Charles et al. 2008 for a recent review). These ``super-orbital'' periods are thought to be related to the properties of the accretion disc, and there are two basic mechanisms actively under consideration: radiation-induced warping (Ogilvie \& Dubus 2001) and precession (Whitehurst \& King 1991) of the accretion disc. Radiation pressure from the intense X-rays arising near the compact object causes the warping of the accretion disc, while tidal forces in high mass-ratio binaries lead to the precession of the accretion disc (an effect first observed in CVs, see e.g. Warner 1995). Either of these effects can lead to the periodic obscuration of the compact object in X-ray binaries, causing a super-orbital modulation in the X-ray flux (Clarkson et al. 2003). The stability criteria established by Ogilvie \& Dubus (2001) suggest that this effect should not arise in most LMXBs, however it is observed, for example, in the long-period LMXB CygX-2 (Clarkson et al. 2003).

The presence of even longer-term quasi-periodic variations (on the order of decades) has been predicted in CVs as a consequence of the modulation in the mass transfer-rate due to a solar-type magnetic-activity cycle occurring in the donor star (Applegate \& Patterson 1987, Warner 1988). The donor stars in short-period LMXBs and CVs are tidally locked and therefore rotate at the orbital period of the binary system, which is $\lesssim1$ d for most systems. This rapid rotation is expected to generate much stronger magnetic fields (Schrijver \& Zwaan 1992). Long-term variability changes in the surface activity of stars (such as starspot activity) have been detected on timescales of decades (Baliunas 1985), similar to the $\sim$11 yr magnetic-activity cycle of our Sun.

Applegate (1992) suggested that magnetically active donors become more oblate as their outer layers are spun up, due to angular momentum distribution changes brought about by their magnetic activity. As a result, the volume of the Roche lobe changes during the magnetic cycle, while the volume of the donor remains unchanged. The magnetic activity cycle therefore governs the Roche-lobe volume and the structure of the donor, which will be varying in oblateness as the cycle progresses. Such a variation will modulate the amount by which the donor overfills the Roche lobe and therefore the mass transfer rate will also vary on the magnetic-activity time-scale (as discussed by Richman, Applegate \& Patterson 1994 for CVs, hereafter R94). 

Considering the similarities between LMXBs and CVs, it is not unreasonable to expect similar longer-term modulations in LMXBs for the same reason. Online archival datasets now allow for a detailed investigation of long-term variations in the lightcurves of X-ray sources. Here we report on an analysis of the $\sim$13-year X-ray history of LMXBs contained in the RXTE/ASM datasets. 

Shortly before completing this paper, we became aware of Durant et al (2009). While they mention similar long-term timescales of the X-ray flux modulations in the 16 brightest persistent LMXBs contained in the RXTE ASM datasets, we propose here a mechanism which would explain the observed dichotomy between Atoll and Z sources, and for the origin of these long time-scales. 

\section{Observations and Data Analysis}

\subsection{RXTE/ASM}

The All Sky Monitor (ASM) aboard the Rossi X-ray Timing Explorer (RXTE) is provided and operated by the Massachusetts Institute of Technology (MIT). The ASM has observed the X-ray sky since early 1996 by scanning approximately 80\% of the sky during each orbit of the satellite, providing monitoring of a source every $\sim$90 minutes for at least 90 seconds. The ASM contains three rotating Scanning Shadow Cameras (SSC), allowing positional measurement of previously unknown sources to within 3 arcmin precision and providing individual monitoring sessions, which are referred to as ``dwells''. 

Data are available in four energy bands: 1.5-3 keV (A), 3-5 keV (B), 5-12 keV (C) and a Sum-band 1.5-12 keV (A+B+C). Data for each energy band are reduced independently by taking background effects into account and made available as dwell-by-dwell or one-day-averages. Data are reduced and compiled weekly by the ASM team and made publically available \footnote{http://xte.mit.edu/ASMlc.html}.  For full details see Levine et al. (1996).

The ASM monitoring began on 20 February 1996, and for this paper we were able to employ datasets spanning more than 13 years.  Therefore these archival datasets contain the long-term intensity history of all X-ray sources in the RXTE catalogue. The fact that RXTE is sensitive in the soft X-ray regime (2-10 keV) is particularly useful for the studies of LMXBs, since their largest flux contribution comes from soft X-rays.

\subsection{Data Reduction}

The full ASM lightcurves of all 44 significantly detected LMXBs (average flux $>$ 0.5 counts s$^{-1}$) were considered for further analysis. Transients were excluded (Aql X-1, GROJ1655-40, GX339-4, H1743-322, 4U1543-47, 4U1608-52, 4U1630-47, XTEJ1550-564, XTEJ1701-462 and XTEJ1859+226), since their outburst(s) are the only reason these sources were detected above the $3 \sigma$ level. 4U1820-30 was excluded, since it is almost certainly a triple system (Chou \& Grindlay 2001), which would have additional dynamics compared to the other LMXBs and is therefore not relevant in our analysis here. Cir X-1 was also excluded, since its highly eccentric orbit (Murdin et al. 1980) will introduce large phase-dependent changes in the donor's Roche lobe.

ASM one-day-average sum-band lightcurve data were plotted on scales appropriate for showing long-term behaviour, and from visual inspection of those lightcurves, it became apparent that some sources display large amplitude, very long-term modulations that appear periodic or quasi-periodic in nature. They are : GX3+1, GX9+1, GX9+9, GX354-0, 4U1636-536, 4U1708-40, 4U1735-444, 4U1746-37 and SerX-1. These are all classified as Atoll sources, with the exception of the X-ray burster 4U1708-40. Re-examining all the Atoll sources, there appear to be very long-term modulations in all of them, although to a lesser extent than those sources listed above. Furthermore, all the Z sources display remarkably steady lightcurves over the long-term. 

For the plots, the Liu et al. (2007) classifications for LMXBs were added in square brackets (see the caption of Fig. \ref{lc1} for a description of these classifications). Our subsequent analysis focussed on the Z and Atoll sources, since they appear to represent the two extremes in long-term periodic behaviour. Therefore, the remaining sources (4U1254-69 [B,D,(SB)], 4U1556-60, 4U1624-49 [D], Her X-1 [P,D,E], 4U1708-40 [B], MXB1730-335 [T,G,B,D,R], KS1731-260 [T,B,(SB)], 4U1822-000, 2A1822-371 [P,E], GS1826-238 [T,B], GRS1915+105 [T,D,M,R], 4U1957+11 [U]) were not considered further. 

\subsection{Variability Analysis}

The variability analysis was conducted on ASM dwell-by-dwell data, which were rebinned into 10-day bins, using the prescribed filters$^{1}$ for constructing ASM averages. These are: $\chi_{\nu}^2<1.5$ ($<8$ for ScoX-1), number of sources in the field of view $<16$, Earth angle $>75$\deg, exposure time $>30$ seconds, long-axis angle: $-41.5$\deg$< \theta <46$ \deg, short-axis angle: $-5$\deg$< \phi <5$\deg. Additional filters applied are: background counts $<10$, hardness ratio: $-5< \frac{B+C}{A}<5$, flux error $<3$, number of datapoints per bin $>10$ to give the most reliable dataset. We include data up to 13 August 2009 ($50100<$ MJD $<55056$).

\begin{figure}
  \centering
	\includegraphics[angle=-90,width=0.47\textwidth]{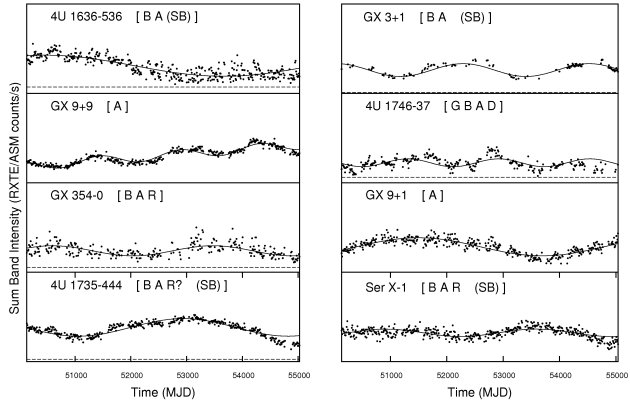}
  \caption {Significantly modulating Atoll sources : Binned 10-day-average RXTE/ASM lightcurves. Indicators in square brackets have the following interpretation: [A] Atoll, [B] X-ray burst, [D] dipping LMXB, [E] eclipsing or partially eclipsing LMXB, [G] globular-cluster, [M] microquasar, [P] X-ray pulsar, [R] radio loud X-ray binary, [T] transient, [U] ultra-soft X-ray spectrum, [Z] Z-type and [(SB)] Super-bursters.} 
  \label{lc1}
\end{figure}

\begin{figure}
  \centering
	\includegraphics[angle=-90,width=0.47\textwidth]{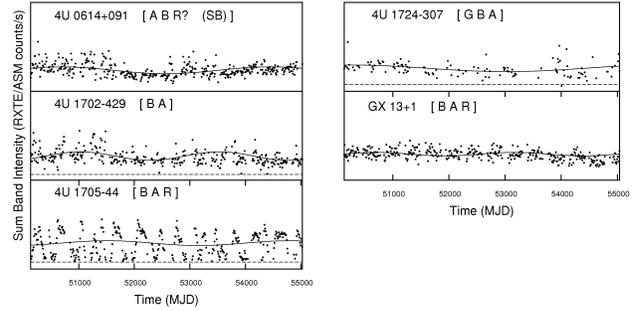}
  \caption {Remaining Atoll sources : Binned 10-day-average RXTE/ASM lightcurves} 
  \label{lc2}
\end{figure}

Since the long-term modulations displayed in these lightcurves have timescales that exceed or are comparable to the observational baseline, we cannot make use of the usual period analysis tools such as periodograms. In order to estimate the time-scale of these long-term variations, single sine-waves were fitted to all the Atoll sources. In the case of GX9+9 a linear term was added (cf. Harris et al. 2009, who also noted the presence of the $\sim$ 1500 d modulation). The results for the 8 Atoll sources which display the most significant long-term modulations, are contained in Table \ref{compareA1} (and their lightcurves in Fig. \ref{lc1}). The results for the remaining Atoll sources are included in Table \ref{compareA2} (and Fig. \ref{lc2}).

\begin{figure}
  \centering
	\includegraphics[angle=-90,width=0.47\textwidth]{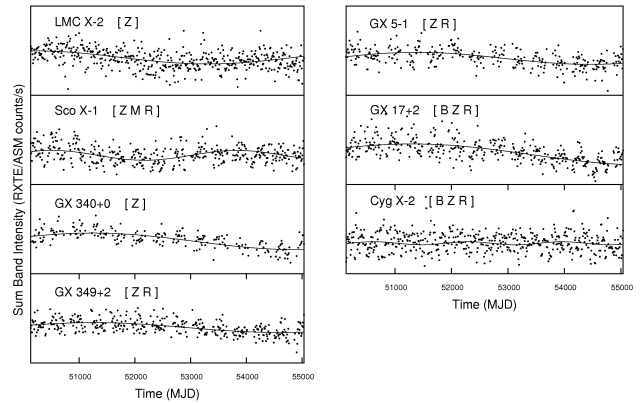}
  \caption {Z sources : Binned 10-day-average RXTE/ASM lightcurves.} 
  \label{lc3}
\end{figure}

Although the Z sources appear to display remarkably steady lightcurves over the long-term, they can also be fitted with single sine-waves, but with much lower amplitudes. The results of their fits are contained in Table \ref{compareZ} and Fig. \ref{lc3}.

\begin{center}
	\begin{table}
	\caption{ASM properties of the 8 significantly modulating Atoll sources}
	\label{compareA1}
	\smallskip
	\begin{tabular}{lccccc}
	\hline
	\bf{Source} & \bf{Average} $^{2}$& \bf{Amplitude} $^{2}$& \bf{Period} $^{2}$& \bf{F}\\
	  & \bf{Flux} &  &  & Stat.\\
	  & \bf{[counts/s]} & \bf{[\% of Flux]} & \bf{[yr]} & \\
	\hline	
	4U 1636-536 & 10.2(2) & 49(4) & 17.6(7) & 336\\
	GX 9+9 & 20(4) & 8(2)  & 3.95(3) & 211\\
	GX 354-0 & 6.8(1) & 31(7) & 7.9(2) & 49\\
	4U 1735-444 & 14.14(7) & 28(1) & 10.3(1) & 491\\
	GX 3+1 & 20.8(2) & 29(4) & 6.07(7) & 133\\
	4U 1746-37 & 2.70(5) & 27(9) & 4.36(8) & 35\\
	GX 9+1 & 38.0(1) & 12.2(5) & 12.0(2) & 391\\
	Ser X-1 & 16.27(8) & 9(2) & 7.3(2) & 58\\
	\hline
	\end{tabular}	
	\end{table}
\end{center}

\begin{center}
	\begin{table}
	\caption{ASM properties of the remaining Atoll sources}
	\label{compareA2}
	\smallskip
	\begin{tabular}{lccccc}
	\hline
	\bf{Source} & \bf{Average} $^{2}$& \bf{Amplitude} $^{2}$& \bf{Period} $^{2}$& \bf{F}\\
	  & \bf{Flux} &  &  & Stat.\\
	  & \bf{[counts/s]} & \bf{[\% of Flux]} & \bf{[yr]} & \\
	\hline
	4U 0614+091 & 3.03(5) & 16(7) & 10.4(7) & 26\\	
	4U 1702-429 & 3.23(7) & 23(11) & 6.9(3) & 19\\
	4U 1705-44 & 13.3(4) & 13(10) & 9(1) & 3\\
	4U 1724-307 & 2.4(1) & 21(28) & 12(3) & 6 \\
	GX 13+1 & 22.61(7) & 2(1) & 6.7(4) & 10 \\
	\hline
	\end{tabular}	
	\end{table}
\end{center}

\begin{center}
	\begin{table}
	\caption{ASM properties of the Z sources}
	\label{compareZ}
	\smallskip
	\begin{tabular}{lccccc}
	\hline
	\bf{Source} & \bf{Average} $^{2}$& \bf{Amplitude} $^{2}$& \bf{Period} $^{2}$ & \bf{F}\\
	  & \bf{Flux} &  &  &  Stat.\\
	  & \bf{[counts/s]} & \bf{[\% of Flux]} & \bf{[yr]} & \\
	\hline
	LMC X-2 & 1.56(2)  & 12(7) & 17(2) & 67\\
	Sco X-1 & 896(3) & 4(1) & 9.1(4) & 27\\
	GX 340+0 & 27.7(3) & 12(8) & 20(2) & 75\\
	GX 349+2 & 49.8(3) & 6(5) & 18(2) & 53\\
	GX 5-1 & 70.7(3) & 4.8(4) & 17(2) & 46\\
	GX 17+2 & 43.4(6) & 7(5) & 23(4) & 97\\
	Cyg X-2 & 37.2(3) & 3(2) & 4.6(4) & 2\\
	\hline
	\end{tabular}	
	\end{table}
\end{center}

Uncertainties in the values obtained are included in the tables in parentheses \footnote{the numbers in parentheses are 1-sigma errors on the fits as described in the text, and refer to the last decimal place quoted}. The binned data flux errors were adjusted by a factor, so as to obtain fits for which $\chi_{\nu}^2\sim1$. This gives more sensible errors on the sine-wave parameters, since the errorbars for the binned data are extremely small in comparison to the larger flux variations (which dominate). The factors applied, to the sources that show the largest scale systematic variations, were: 43.4 for Cyg X-2, 36.8 for 4U 1705-44 and 32.9 for Sco X-1. The factors applied to the rest of the sources range from 1.5 - 14.7, which are comparable to the factor of 3.1 required for a constant fit to the Crab.

The F-statistic was calculated for the 2 models of a constant fit and a single sine-wave fit to the data. For GX9+9 the simpler model considered was a linear variation with time. The calculated F-values are included in Tables \ref{compareA1}, \ref{compareA2} \& \ref{compareZ}. F-values $>>1$ are obtained, indicating that the inclusion of additional terms (the sine wave) is highly significant ($>99\%$ confidence). The F-values are highest in those sources where we could detect long-term variation by simple visual inspection. Even the lowest F-values determined $\sim(2-10)$ are still highly significant, indicating that all the sources considered here were better fitted with a single sine-wave than with a constant value.

Flux modulations of $\sim8-30\%$ of the average flux values are present in the Atoll sources, with the exception of the superburster 4U1636-536 and the burster GX13+1. It is possible that the modulation detected in 4U1636-536 represents a superburst, rather than the long-term modulations we consider to be present in the other sources (which may also be the case for KS1731-260). GX13+1 is classified as an Atoll source, but it shares certain properties with Z sources (Liu et al. 2007). Indeed, we find that its amplitude for the very long-term modulation rather agrees with those found in the Z sources, than with those obtained for the Atoll sources. Flux modulations are $<$13\% for the Z sources, with the brighter sources having the lower amplitudes.

\section{Discussion}

Both Atoll and Z sources contain NS primaries, but Z sources have fluxes that are $\sim$0.5-1$L_{Edd}$, whereas Atoll sources and bursters have fluxes $\sim$0.01-0.5$L_{Edd}$ (van der Klis 2006). Once at the Eddington limit ($L_{Edd}$), Z sources are unlikely to show any X-ray flux modulation due to additional changes in the mass transfer rate. However, Atoll sources would be expected to modulate their flux in response to overall changes in the mass transfer rate. The results show that, in general, Atoll sources have larger amplitude in the very long-term modulations than Z sources, which occur on very long-term timescales in both types. 

\subsection{Solar-cycle type timescales in LMXBs?}

A mechanism for modulating the mass transfer rate over long timescales has been proposed for CVs (Applegate \& Patterson 1987, Warner 1988). Given the similarities between CVs and LMXBs, we decided to investigate whether this mechanism might also be applicable to LMXBs.

Our current X-ray observational baseline for LMXBs is too short to cover multiple cycles and thereby establish the stability of these variations. However, given the similarity of these timescales to those exhibited by solar-type stars, we use the approach of R94 for CVs, and see if it is also applicable to LMXBs. 

The mechanism proposed in R94 to be responsible for long-term variations in CVs is due to magnetic activity cycles in the donors. They calculate the variation in the mass transfer rate ($\frac{\Delta \dot M}{\dot M}$) which is associated with the observed orbital period variation ($\frac{\Delta P}{P}$) brought about by this mechanism in CVs.

R94 proposed that changes in the rotation of a thin outer shell (mass $M_s$) of the donor (mass $M_{2}$), rotating with angular velocity ($\Omega$), will affect the orbital period. They calculate that:
\begin{equation}
\frac{\Delta P}{P} = -0.04 \left(\frac{q}{1+q}\right)^{2/3} \frac{M_s}{M_2} \frac{\Delta \Omega}{\Omega}
\end{equation}
where $q = \frac{M_2}{M_1}$. They noted that the Applegate (1992) variable differential rotation rates follow the Hall (1990) and Hall (1991) differential rotation-orbital period relation, and consequently applied that relation to the orbital periods for CVs to obtain $\frac{\Delta \Omega}{\Omega} \sim 0.0015 $. They furthermore assume that $\frac{M_s}{M_2} \sim 0.1$ and calculate a $\frac{\Delta \dot M}{\dot M}$ which is consistent with the observed long-term flux variations in CVs, but consider the observed $\frac{\Delta P}{P}$ to be the best evidence for long-term cycles with a preferred timescale of decades (5-30 yrs), reminiscent of solar-like magnetic cycles (Warner 1988). They also find that CVs show quasi-periodic brightness fluctuations over that timescale.

We apply equation (1) to the LMXB GX9+9, which has $P_{orb} = 4.1958 \pm 0.0005$ hrs (Kong et al. 2006), and for which $q = 0.25$ has been found spectroscopically by Cornelisse et al. (2007). We consequently obtain $\frac{\Delta P}{P} = -2\times10^{-6} $. 

Cornelisse et al. (2008) found no significant change in orbital period over the $\sim$11 yr RXTE/ASM baseline. Considering the result for $\frac{\Delta P}{P}$, we would not expect to detect a change in the orbital period in the RXTE/ASM dwell-by-dwell data, since the $\Delta P$ implied is $\sim$ 60 times smaller than the error in $P_{orb}$.

Such a change in P will cause a corresponding change in the size of the donor's Roche lobe ($R_2$) and therefore modulate the mass transfer rate $\dot M$ by an amount determined by R94
\begin{equation}
\frac{\Delta \dot M}{\dot M} = -\frac{1}{3} \left(\frac{a}{R_2}\right)^2 \frac{R_2}{H} \frac{M_2}{M_s} \frac{\Delta P}{P}
\end{equation}
where $H$ is the photospheric scale-height of a main sequence donor. Assuming the standard Paczy\'{n}ski (1971) relation for the size of the donor's Roche lobe gives $\frac{R_2}{a} = 0.27$ and if the donor in GX9+9 also follows the CV secondary relation of $ \frac{R}{R_{\odot}} = \left( \frac{M}{M_{\odot}} \right) ^{0.88} $ , then $ \frac{\Delta \dot M}{\dot M} =  0.3 $ for $ \frac{R_2}{H} = 3200 $ (see R94).

Therefore a maximum flux modulation of $\sim30\%$ is expected by virtue of the orbital period variation and we actually observe $\sim$ 8 $\%$ over a period $P_{long} \sim 4 $ yrs. In fact, this result regarding the expected maximum flux modulation will apply to all the LMXBs mentioned in this paper, for the whole range of long-term periods determined from the fitted sine-waves. The flux modulations determined from the fitted sine-waves for the sources considered are $\lesssim30\%$. It is therefore quite plausible that the long-term modulations observed in the RXTE ASM lightcurves of the LMXBs, originate from magnetic activity cycles in the donor, just as has been proposed for CVs.

\section{Conclusions}

Long-term variation in the mass transfer rate due to the magnetic activity cycle of the donor, should therefore translate into long-term quasi-periodic modulations of the X-ray flux for sources in which the additional material can be accreted onto the NS without violating the Eddington limit, such as Atoll sources and bursters. However, in Z sources very little (if any) of the additional material will be accreted and much lower amplitude (if any) long-term X-ray flux modulations are expected as a result of the magnetic activity cycle of the donor.

Therefore, we conclude that RXTE ASM lightcurve data may now provide the first evidence for very long-term quasi-periodic modulations of the X-ray flux as a result of the modulation in the mass transfer-rate due to a solar-type magnetic-activity cycle in the donor star, similar to those proposed for CVs.

\section{Acknowledgements}
ASM results were provided by the ASM/RXTE teams at MIT and at the RXTE Science Operations Facility and Guest Observer Facility at NASA's Goddard Space Flight Center (GSFC). We also wish to thank Brian Warner for his support and comments.

\label{lastpage}

\end{document}